\documentclass[12pt]{article}
\usepackage[dvips]{graphicx}

\usepackage{graphicx}
\usepackage{graphics}
\usepackage{epsfig}
\usepackage{dcolumn}
\usepackage{bm}

\def\be{\begin{equation}}
\def\ee{\end{equation}}
\def\bdm{\begin{displaymath}}
\def\edm{\end{displaymath}}

\begin{document}
\title{Macroscopic description for a quantum plasma micro-instability: the quantum Weibel solution}
\author{F. Haas \footnote{Also at Universidade do Vale do Rio dos Sinos - UNISINOS, Av. Unisinos 950, 93022--000, S\~ao Leopoldo, RS, Brazil} \,\, and M. Lazar}
\date{\relax}
\maketitle
\begin{center}
{Institut f\"ur Theoretische Physik IV, 
Ruhr-Universit\"at Bochum\\ D-44780 Bochum Germany}
\end{center}

\begin{abstract}
The Weibel instability in the quantum plasma case is treated by means of a fluid-like (moments) approach. Quantum modifications to the macroscopic equations are then identified as effects of first or second kind. Quantum effects of the first kind correspond to a dispersive term, similar to the Bohm potential in the quantum hydrodynamic equations for plasmas. Effects of the second kind are due to the Fermi statistics of the charge carriers and can become the dominant influence for strong  degeneracy. The macroscopic dispersion relations are of higher order than those for the classical Weibel instability. This corresponds to the presence of a cutoff wave-number even for the strong temperature anisotropy case. 
\end{abstract}


\maketitle

\section{Introduction}

The field of quantum plasmas has been introduced long ago \cite{klimontovich, pines} and is  presently attracting renewed attention from a variety of viewpoints. It was already confirmed that quantum mechanical effects, e.g., electron tunneling and wave-packet spreading, play a central r\^ole in the behavior of metallic or semiconductor nanostructures of the next generation electronic devices \cite{gr96}--\cite{a07}. Some astrophysical compact objects, such as white dwarf or neutron stars, possess very high temperature but strong quantum effects as well due to their large densities ($\sim 10^6$ g/cm$^3$) \cite{s83}. There has been recent studies in quantum plasmas involving quantum turbulence \cite{shu1}, quantum analogues for the Harris sheet \cite{h}, quantum models taking into account spin \cite{m1, m2}, stable solitary structures \cite{be}, dark soliton and vortices solutions \cite{pr}, variational structures for the quantum Zakharov system \cite{hz} as well as application of quantum hydrodynamic equations for carbon nanotubes \cite{wei}. 

The growing interest on quantum plasmas comes in part from the recently introduced hydrodynamic equations \cite{haas1}-\cite{haas2}, which are simpler in comparison to the kinetic descriptions used in the original developments. However, the Weibel instability \cite{weibel} is usually treated in terms of kinetic descriptions. The Weibel instability is one of the basic plasma instabilities and is driven by an anisotropic velocity distribution of plasma particles \cite{weibel, f59}. The quantum version of the Weibel instability has been recently proposed \cite{haas3, levan} on grounds of the dispersion relation for the Wigner-Maxwell system, which is the quantum counterpart of the Vlasov-Maxwell system. Therefore, the details of the instability are dependent on the precise form of the equilibrium Wigner pseudo distribution function, in a similar way as the traditional Weibel instability is partially dependent on the exact form of the classical equilibrium distribution function. The purpose of this paper is to overcome this difficulty by means of a moment description for the quantum Weibel instability. Recently, the classical Weibel instability was investigated by Basu \cite{basu} taking moments of the Vlasov-Poisson system and the present work follows basically the same strategy.  Here, however, the starting point is the linearized Wigner-Maxwell system. It is also interesting to verify to what extent a fluid-like approach as the moment method is able to capture the essentials of the Weibel instability, in the quantum case. Some peculiar subtleties coming from the quantum nature of the model equations will show up.  The transition from a kinetic to a fluid-like approach in a quantum plasma model will be shown to give rise to quantum effects of a different nature according to the density of the system, as  explained more thoroughly in the continuation. 

Classical plasmas frequently have equilibrium distribution functions a\-ni\-so\-tro\-pic in velocity space \cite{r5a}-\cite{r7}. In the context of quantum plasmas, velocity anisotropy can arises at least for laser plasmas and neutron stars. It is well-known \cite{Estabrook78} that anisotropic heating by resonant absorption can produce a Weibel-like instability in laser plasmas. Also, there are experimental evidence of Weibel instability in laser-solid interaction experiments \cite{Wei04}. In addition, in tunnel-ionized laser plasmas there can be velocity anisotropy driven by a varying laser polarization \cite{Leemans}. Quantum effects should be more evident in the next generation of laser-solid interaction experiments, where the densities are very high. For neutron stars, it has been conjectured \cite{Wanajo2006} that anisotropic heating can arise in view of fast rotation, implying a strongly deformed neutrino sphere and anisotropic neutrino fluxes. There are estimates \cite{Walder2005} where the pole-to-equator neutrino flux ratio can assume a value of 2. For these reasons, it is important to have a better understanding of the Weibel instability taking into account quantum effects. 

As examples of distinct equilibrium Wigner functions for the quantum Weibel instability, one can have Maxwell-Boltzmann or Fermi-Dirac functions, both with anisotropy in velocity space. Using a moment description, there is some lost of information, but more universal statements are made available. As for any moments or fluid modeling, an intrinsic limit of such approach is in the closure of the equations. Indeed, one is always faced with a system where the equation for the time evolution of the velocity moment of order $n$ depends on the velocity moment of order $n+1$. In this way \cite{basu}, it happens that the moment approach is appropriate only for long wave-length and large temperature anisotropy. Moment descriptions have also been applied to cyclotron wave-particle interaction \cite{siregar}. 

This work is organized as follows. In Sec. II we construct the general formalism 
of the moment equations using the linearized Wigner equation. Assuming a large 
temperature anisotropy, we derive the new dispersion relation for the electromagnetic unstable modes of Weibel-type, which includes quantum corrections appropriate for dilute systems. In Sec. III we generalize the analysis for an anisotropic Fermi-Dirac distribution. The fourth-order moment term provides in this case a quantum correction term of a different nature in the dispersion relation. Our quantum dispersion relations are discussed in Sec. IV. The analytical forms for the Weibel growth rates are derived and plotted for representative highly dense plasmas. A brief summary of the results is given in Sec. V.

\section{Basic equations}
Consider a quantum plasma with equilibrium Wigner function $f = f_{0}({\bf v})$ and no equilibrium electromagnetic field. If $\tilde{f} = \tilde{f}({\bf r}, {\bf v}, t)$ and $\tilde{\bf A} = \tilde{\bf A}({\bf r}, {\bf v}, t)$  denotes the perturbations of the equilibrium Wigner function and of the vector potential, then the linearized Wigner equation \cite{haas3} reads
\begin{eqnarray}
\label{e1}
\frac{\partial\tilde{f}}{\partial t} &+& v_{i}\left(\frac{\partial\tilde{f}}{\partial r_i} + \frac{e}{m}\frac{\partial\tilde{A}_j}{\partial r_i}\frac{\partial f_0}{\partial v_{j}}\right) + \frac{e}{m}\frac{\partial\tilde{\bf A}}{\partial t}\cdot\frac{\partial f_0}{\partial{\bf v}}   \\  &-& \frac{ie}{\hbar}\left(\frac{m}{2\pi\hbar}\right)^3 {\bf v}\cdot\int d{\bf s} d{\bf v}' e^{im({\bf v} - {\bf v}')\cdot{\bf s}/\hbar} \, [\tilde{\bf A}({\bf r} + \frac{\bf s}{2}) - \tilde{\bf A}({\bf r} - \frac{\bf s}{2}) ] f_{0}({\bf v}') = 0 \,. \nonumber
\end{eqnarray}
In the above equation, the summation convention is used in some terms and it is assumed the Coulomb gauge $\nabla\cdot\tilde{\bf A} = 0$, as well as the perturbed electrostatic potential is taken to be zero. In addition, $\hbar = h/(2\pi)$ is the scaled Planck constant, $-e$ is the electron charge and $m$ the electron mass. Furthermore, the treatment is restricted to transverse waves so that 
\begin{equation}
\label{e2}
\tilde{\bf A} = {\bf A}_{\bot} \exp(i[kz-\omega\,t]) \,,
\end{equation}
where ${\bf k} = k\hat{z}$ is the wave vector and ${\bf A}_{\bot}$ is a constant vector satisfying ${\bf k}\cdot {\bf A}_{\bot} = 0$. In all calculations, $\partial/\partial z$ is the only spatial derivative which does not identically vanishes. Also, the equilibrium Wigner function is an even function of the velocity components. 

It is convenient to define the first, second and third order moments
\begin{eqnarray}
\label{e3}
\tilde{u}_x &=& \frac{1}{n_0} \int d{\bf v} v_x \tilde{f}({\bf r}, {\bf v}, t)  \,,\\
\label{e4}
\tilde{P}_{xz} &=& m \int d{\bf v} v_x v_z \tilde{f}({\bf r}, {\bf v}, t)  \,,\\
\label{e5}
\tilde{Q}_{xzz} &=& m \int d{\bf v} v_x v_{z}^2 \tilde{f}({\bf r}, {\bf v}, t)  \,,
\end{eqnarray}
where $n_0 =  \int d{\bf v} f_{0}({\bf v})$ is the equilibrium density. From (\ref{e1}), after integrating by parts and taking into account the Coulomb gauge as well as the parity properties of $f_0$, one get
\begin{eqnarray}
\label{e6}
\frac{\partial}{\partial t}\tilde{u}_x &=& - \frac{1}{mn_0} \frac{\partial}{\partial z}\tilde{P}_{xz} - \frac{e}{m}\tilde{E}_x \,, \\
\label{e7} 
\frac{\partial}{\partial t}\tilde{P}_{xz} &+& \frac{\partial}{\partial z}\tilde{Q}_{xzz} = \frac{e n_0}{m}\tilde{B}_{y} (T_{\parallel} - T_{\bot})   \,, \\
\label{e8}
\frac{\partial}{\partial t}\tilde{Q}_{xzz} &+& m\frac{\partial}{\partial z}\int d{\bf v} v_{x} v_{z}^3 \tilde{f}({\bf r},{\bf v},t) = - \frac{e n_{0} T_{\parallel}}{m} \tilde{E}_{x}     \,,
\end{eqnarray}
which are exactly the same as Eqs. (10-12) from Basu's work \cite{basu}, in a different notation. In (\ref{e6}-\ref{e8}), $\tilde{E}_x = - \partial\tilde{A}_{x}/\partial t$ and $\tilde{B}_y = \partial\tilde{A}_{x}/\partial z$ are the $x$ and $y$ components of the perturbed electric and magnetic fields, respectively. Also,
\begin{eqnarray}
\label{e9}
T_\parallel &=& (m/n_0) \int d{\bf v} v_{z}^{2} f_{0}({\bf v})  \,,\\
\label{e10}
T_\bot &=& (m/n_0) \int d{\bf v} v_{x}^{2} f_{0}({\bf v})  
\end{eqnarray}
are related to velocity dispersion along the $x$ and $z$ directions, respectively 

The fact that the moment equations following from the (quantum) Wigner equation and the (classical) Vlasov equation are the same seems to be a puzzle. Some quantum contribution should survive, otherwise both classical and quantum dispersion relations would be the same. The key to solve the puzzle is hidden in the 
fourth-order moment term at (\ref{e8}). This term is neglected in the pure classical case, but in the following it is shown that this cannot be taken from granted in the quantum case. 

To estimate the fourth-order moment term at (\ref{e8}), one uses the linearized Wigner equation to find 
\begin{equation}
\label{e11}
\frac{\partial}{\partial t}\int d{\bf v} v_{x} v_{z}^3 \tilde{f}({\bf r},{\bf v},t) = \frac{e}{m}\tilde{B}_{y} I+ \frac{e n_0 \hbar^2 T_\bot}{4m^2} \frac{\partial^2 \tilde{B}_y}{\partial z^2} \,,
\end{equation}
where 
\begin{equation}
\label{e12}
I = \int d{\bf v} (v_{z}^4 - 3 v_{x}^2 v_{y}^2) f_{0}({\bf v}) 
\end{equation}
and the fifth-order moment $\int d{\bf v} v_x v_{z}^4 \tilde{f}$ was disregarded to get closure of the system. In (\ref{e11}), the term proportional to $\hbar^2$ has a quantum nature, while the quantity $I$ can be shown to be negligible in the case of a Maxwell-Boltzmann equilibrium. At this point,  suppose that $I$ produces only a higher-order correction, an approximation to be checked in more detail in Section III. Assuming $I \approx 0$ and Fourier transforming with all quantities proportional to $\exp(i[kz-\omega t])$ in (\ref{e6}-\ref{e8}), (\ref{e11}) and in Faraday and Amp\`ere laws, there follows the dispersion relation
\begin{equation}
\label{e13}
\omega^2 - c^2 k^2 - \omega_{p}^2 \left[1 + \frac{k^2 T_\bot}{m\omega^2} \left(1 + \frac{\hbar^2 k^4}{4m^2 \omega^2} \right)\right] = 0 \,,
\end{equation}
where $\omega_p = (n_0 e^{2}/(m\varepsilon_0))^{1/2}$ is the plasma frequency and $c$ the speed of light. Equation (\ref{e13}) is the same as Eq. (22) of Basu's work \cite{basu}, but now with the extra quantum term proportional to $\hbar^2$. Notice that the final result is independent of $T_\parallel$, a fact which is consistent with an extreme temperature anisotropy assumption ($T_\bot \gg T_\parallel$). 

Until now the treatment is completely general, with no particular assumption on the form of the equilibrium distribution function, as long as $I$ in (\ref{e12}) can be disregarded. In this sense, the instability follows from temperature anisotropy,  whatever the exact form of the equilibrium distribution function. Nevertheless, in Section III it is shown that for extreme degenerate Fermi gases one is obliged to fully keep the fourth-order moment contribution, including the term $I$ which would be not negligible anymore. This leads to a modified dispersion relation, useful for very dense plasmas like in astrophysical objects as white dwarfs and neutron stars as well as in laser-solid plasma interaction experiments. It can be said, that the quantum correction in the second term at the right-hand side of (\ref{e11}) is always present and that an additional quantum correction coming from extreme densities can also manifest through the term $I$, fairly negligible for classical plasma. Modifications arising from the dispersive term $\sim \hbar^2$ at (\ref{e11}) will be referred in the present context as (quantum) effects of the first kind, while the contribution from the $I$ integral will be called a perturbation of the second kind. 

In order to compare the dispersion relation (\ref{e13}) to previous work on the quantum Weibel instability, one can insert the extreme anisotropic equilibrium distribution function
\begin{equation}
\label{e14}
f_0 = \frac{n_0 m}{2\pi T_\bot} \delta(v_z) \exp\left[-\frac{m}{2T_{\bot}} (v_{x}^2 + v_{y}^2)\right]
\end{equation}
into the Wigner-Maxwell system as it is presented, for instance, in reference \cite{haas3}. After linearizing and Fourier transforming, the result is 
\begin{equation}
\label{e15}
\omega^2 - c^2 k^2 - \omega_{p}^2 \left[1 + \frac{k^2 T_\bot}{m\omega^2} \left(1 - \frac{\hbar^2 k^4}{4m^2 \omega^2}\right)^{-1}\right] = 0 \,,
\end{equation}
which is the same as (\ref{e13}) provided $\hbar^2 k^4/(4 m^2 \omega^2) \ll 1$, in accordance with the long wave-length approximation. If one proceeds with (\ref{e15}), one would also get the dispersion relation shown in Eq. (29) of reference \cite{haas3}. Hence, the moment  and the kinetic theory approaches gives the same results, provided there is sufficient temperature anisotropy and the long wave-length assumption is valid. 

\section{Anisotropic Fermi-Dirac equilibrium} 

It should be observed that quantum effects in plasma can be taken into account in at least two ways.  On one hand, a quantum transport equation can be the starting point. In this work, the r\^ole of quantum transport equation is played by the linearized Wigner equation (\ref{e1}). Unlike Vlasov's equation, the Wigner equation is able to model quantum phenomena like tunneling and wave-packet spreading. On the other hand, quantum effects can be incorporated by means of an equilibrium distribution reflecting the spin of the charge carriers. This second avenue is pursued in this Section, where radical departures to the dispersion relation are found, especially for strongly degenerate systems. 

In the previous Section, the quantity $I$ at (\ref{e12}) was neglected and the dispersion relation (\ref{e13}) was obtained. The purpose of this Section is to investigate more closely the assumption on the smallness of $I$. In order to get closure of the moment equations, it is unavoidable to add some hypothesis on the equilibria. As will be shown, it is not generically true that $I$ can be neglected. Indeed, one can consider the equilibrium Wigner function appropriate for an anisotropic Fermi-Dirac distribution, 
\begin{equation}
\label{e16}
f_0 = \frac{\alpha}{\exp\left[\frac{m}{2}\left(\frac{v_{x}^2 + v_{y}^2}{T_{\bot}} 
+ \frac{v_{z}^2}{T_{\parallel}}\right) - \beta\mu\right] + 1} \,, 
\end{equation}
where $\mu$ is the chemical potential and $\alpha$ is a normalization constant, 
\begin{equation}
\label{e17}
\alpha = - \frac{n_{0}}{{\rm Li}_{3/2}(- e^{\beta\mu})} \Bigl(\frac{m\beta}{2\pi}\Bigr)^{3/2} 
= 2\Bigl(\frac{m}{2\pi\hbar}\Bigr)^3   \,.
\end{equation}
In (\ref{e17}), ${\rm Li}_{3/2}$ is a polylogarithm function \cite{lewin}. In addition,  $\beta = 1/[(T_{\bot}^2 T_{\parallel})^{1/3}]$, with the temperatures $T_\bot$ and $T_\parallel$ measured in terms of the Boltzmann constant. If $T_\bot = T_\parallel$, the standard Fermi-Dirac statistics is recovered. The Fermi statistics is unavoidable in the case of degenerate Fermi gases, as intense laser beams or compact astrophysical objects. Dilute systems ($e^{\beta\mu} \ll 1$) are fairly well treated by the Maxwell-Boltzmann equilibrium. Notice that (\ref{e16}) is not the more usual Fermi-Dirac distribution $\hat{f}({\bf k})$, where ${\bf k}$ is the appropriated wave vector in momentum space, but the associated equilibrium Wigner function. These objects are related by $\hat{f}({\bf k}) = (1/2) (2\pi\hbar/m)^3 f_{0}({\bf v})$, with the factor $2$ coming from spin \cite{Ross, Arista}. Another distinctive feature is that here temperature anisotropy is allowed. 

Inserting (\ref{e17}) into (\ref{e12}), the result is 
\begin{equation}
\label{e18}
I = \frac{3 n_0 T_{\parallel} (T_\parallel - T_{\bot})}{m^2} \, \frac{{\rm Li}_{7/2}(- e^{\beta\mu})}{{\rm Li}_{3/2}(- e^{\beta\mu})} \,,
\end{equation}
where ${\rm Li}_{7/2}$ is another polylogarithm function. In the particular case of dilute systems, using the properties of the polylogarithm function , the last equation reduces to 
\begin{equation}
\label{e19}
I = \frac{3 n_0 T_{\parallel} (T_\parallel - T_{\bot})}{m^2}  \,,
\end{equation}
which is equivalent to Eq. (23) of \cite{basu}. In the general case, proceeding as before but retaining the $I$ contribution, there follows the dispersion relation
\begin{equation}
\label{e20}
\omega^2 - c^2 k^2 - \omega_{p}^2 \left\{1 + \frac{k^2 T_\bot}{m\omega^2} \left[1 + \frac     {3k^2 T_{\parallel}(T_\bot - T_{\parallel})}{m\omega^2 T_\bot}\frac{{\rm Li}_{7/2}(- e^{\beta\mu})}{{\rm Li}_{3/2}(- e^{\beta\mu})} + \frac{\hbar^2 k^4}{4m^2 \omega^2}\right] \right\} = 0 \,.
\end{equation}
By inspection, and taken into account the strong anisotropy assumption ($T_\bot \gg T_\parallel$), the dispersion relation (\ref{e20}) is equivalent to the previous one Eq. (\ref{e13}) provided
\begin{equation}
\label{e21}
\frac{\omega^2}{k^2 v_{\parallel}^2} \gg \frac{{\rm Li}_{7/2}(- e^{\beta\mu})}{{\rm Li}_{3/2}(- e^{\beta\mu})} \,,
\end{equation}
where $v_\parallel = (T_{\parallel}/m)^{1/2}$ is the characteristic speed associated to $T_\parallel$. While (\ref{e21}) is automatically satisfied for dilute systems due to the long wave-length approximation, the same is not true for a strongly degenerate Fermi gas. For instance, for $\beta\mu \sim 200$, the right hand side of (\ref{e21}) is as large as $4500$, so that the contribution coming from quantum statistics cannot be neglected at (\ref{e20}). This modified dispersion relation can be useful for a better understanding of the Weibel instability in very dense plasma systems. However, it is a macroscopic relation not so easily comparable to kinetic (Wigner-Maxwell) relations. This is the case, since the anisotropic Fermi-Dirac equilibrium (\ref{e16}) is not easily amenable to analytic results even for extreme temperature anisotropy. However, the analytical difficulties of the kinetic dispersion relation arising from (\ref{e16}) are just one reason more to emphasize the relevance of the macroscopic equation (\ref{e20}).

\section{Numerical solutions and discussions}

To find the Weibel solutions of Eqs. (\ref{e13}) and (\ref{e20})
one should observe that both equations are of third-order in $\omega^2$.
For physically reasonable parameters, there are one real and two complex solutions. As for classical plasma, the complex solutions can be taken in the form of purely growing or evanescent modes, according to $\omega = \imath \Im (\omega) = \imath \gamma$ with $|\gamma|$ usually small, $|\gamma| \ll \omega_p$. Therefore, the first term in both equations (\ref{e13}) and (\ref{e20}) can be neglected for the purpose of calculating the Weibel growth rate.
\begin{figure}[htb]
\centering
\includegraphics[width=7cm]{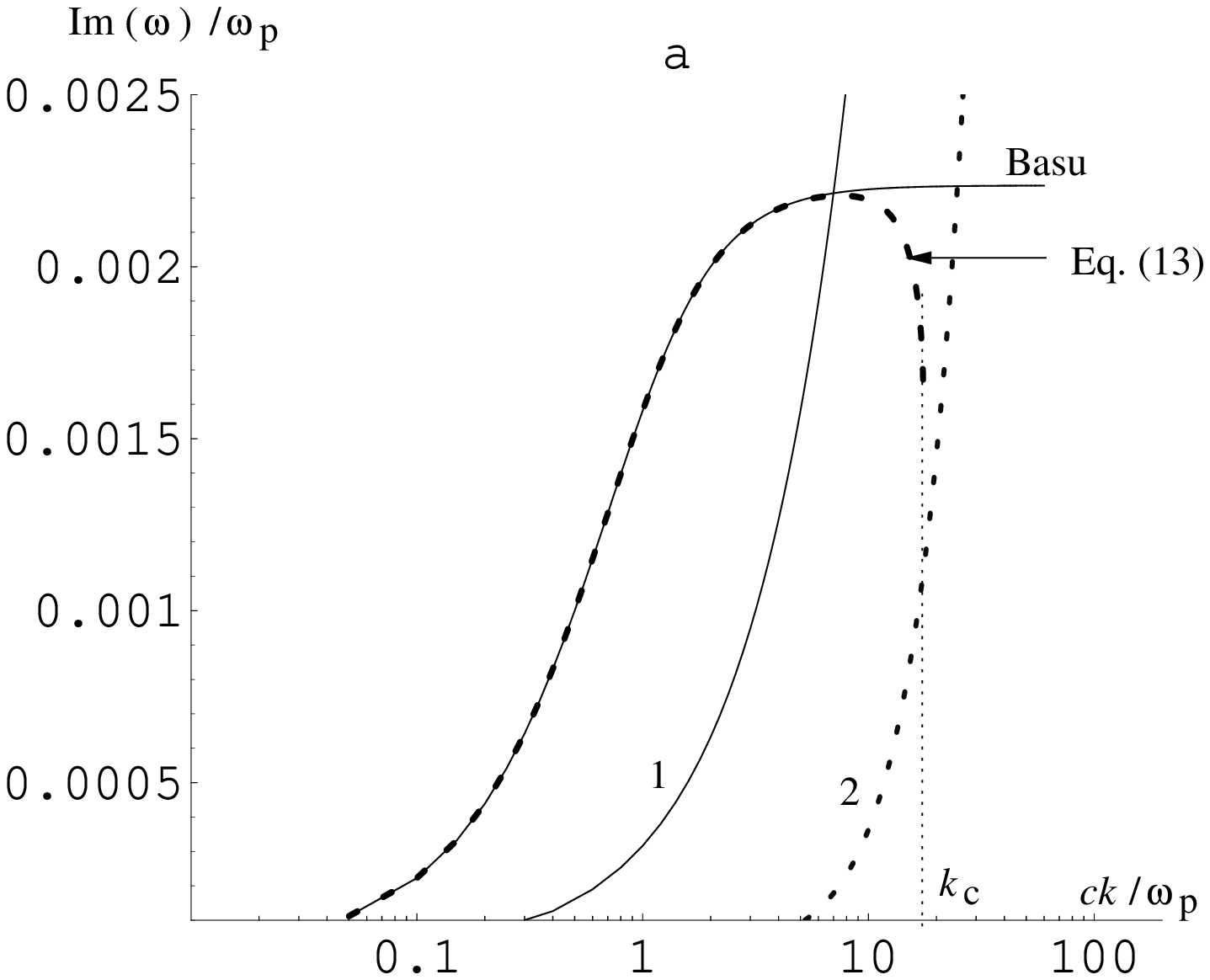} $\;\;\;\;$
\includegraphics[width=7cm]{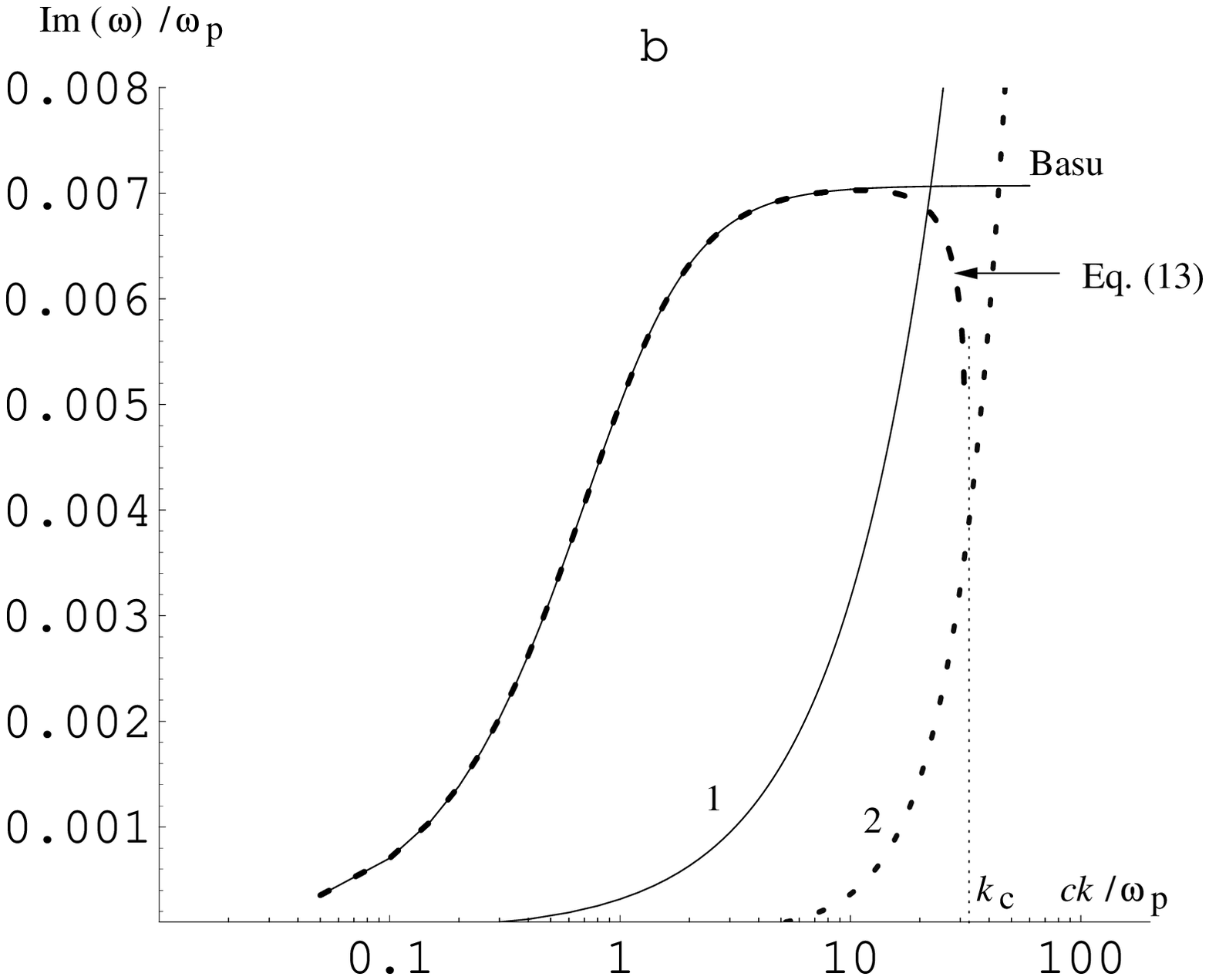}
\caption{With dashed bold lines are shown the Weibel growth rates 
obtained from Eq. (13) for a metallic (gold) plasma 
with $n_0= 10^{28}$ m$^{-3}$ and two temperatures 
(a) $T_{\perp} = 2.5$ eV, and (b) $T_{\perp} = 25$ eV.
In contrast to the classical theory (solid bold lines), the aperiodic solutions of 
Eq. (13) are limited here to wave-numbers $k < k_c$, 
by the quantum effects of the first kind.} \label{fig1}
\end{figure}

Here, one first restrict to the quantum effects described by Eq. (\ref{e13}) 
from which there follows the fourth-order dispersion relation
\begin{equation}
\label{e22}
\left(\frac{c^2 k^2}{\omega_{p}^2} +1 \right)\gamma^4  - 
\frac{k^2 T_\bot}{m} \left(\gamma^2 - \frac{\hbar^2 k^4}{4m^2} \right) = 0 \,,
\end{equation}
which admits four aperiodic solutions given analytically by
\begin{equation}
\label{e23}
\gamma^2 = \frac{k^2 T_\bot}{2 m} \left(\frac{c^2k^2}{\omega_p^2} +1\right)^{-1} 
\left\{1 \pm \left[1- \frac{\hbar^2 k^2 }{m T_\bot}\left(\frac{c^2k^2}{\omega_p^2} +1\right) \right]^{1/2} \right\} \,,
\end{equation}
which exist as long as the wave number is less than a cutoff value, $k \le k_c$.
The cutoff value is given by the existence condition for the square root in (\ref{e23}), 
\begin{equation}
\label{e24}
k_{c}^2 = \frac{\omega_{p}^2}{2c^2} \left[ \left(1+ \frac{4 T_\bot m c^2}
{\hbar^2 \omega_{p}^2}\right)^{1/2} -1 \right] \,.
\end{equation}

For a complete characterization of the Weibel instability, 
we plot the growth rates in Fig. \ref{fig1}. 
It has been chosen a representative 
case of a metallic gold plasma \cite{manfredifields} with density
$n_0= 10^{28}$ m$^{-3}$ and two temperatures 
(a) $T_{\perp} = 2.5$ eV, and (b) $T_{\perp} = 25$ eV. 
One can assume sufficiently large anisotropies, $T_{\perp} / T_{\parallel} =$ 100 $\sim$ 1000,
so that the parallel temperature is close to the room temperature, $T_{\parallel} \simeq 0.025$ eV.

In addition, curve 1 (solid line) is given by the condition 
$|\omega /(kv_{\parallel})| \gg 1$ 
introduced in Basu's Vlasov model  \cite{basu} for the strong temperature anisotropy approximation. 
This curve limits to its left side the existence of the {\it macroscopic} Weibel modes, 
and their fluid approach using Eq. (22) from Ref. \cite{basu}.
On the other hand, curve 2 (dashed line) is given by $\hbar^2 k^4 /(4m^2 \omega^2) \ll 1$ being less restrictive, and limits to its left side the moment description of
the quantum Weibel modes by using Eq. (13). In practice it excludes the lower mode (curve 2 at Figures 1a and 1b as unphysical. In addition, 
their existence is limited only to the wave-numbers 
smaller than a cutoff value, $k < k_c$. This cutoff wave-number is projected with dotted line in Fig. \ref{fig1} (b). It should also be given by the condition 
for a maximum wave-number, $d k/ d\gamma = 0$. Imposing
this condition to the last dispersion relation (\ref{e22}) one find
\begin{equation}
\label{e25}
\gamma_c^2 = \omega_p^2 \frac{k_c^2 T_\bot}{m (c^2 k_c^2 + \omega_p^2)}\,.
\end{equation}
The cutoff wave-number, $k_c$, and the corresponding growth rate 
$\omega_c$ are solutions of Eq. (\ref{e22}), and therefore replacing
(\ref{e25}) in (\ref{e22}) yields exactly (\ref{e24}).
This can be used to evaluate, for example, in Fig. \ref{fig1} (b) the 
cutoff wave-number scaled as $ck_c / \omega_{p} \simeq 6.72$. 
Concluding, Eq. (\ref{e13}) admits four aperiodic solutions 
for each wave-number $k < k_c$.

\begin{figure}[htb]
\centering
\includegraphics[width=6cm]{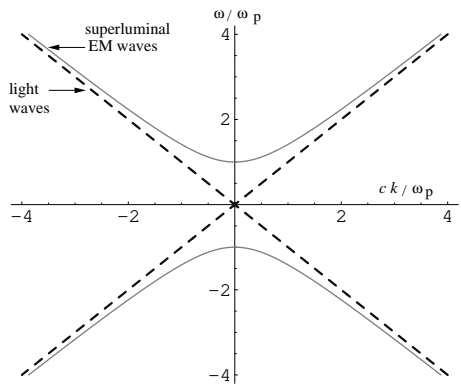}
\caption{With solid lines are shown the superluminal waves 
(no damping and no growing, $\Im (\omega) = 0$) described by Eqs. (\ref{e13}) and 
(\ref{e20}). In this case the wave dispersion is not affected 
by quantum effects. With dashed line are shown the light waves.} \label{fig2}
\end{figure}

The other two solutions of Eq. (\ref{e13}) are real and are plotted in Fig. \ref{fig2}, corresponding to $|\omega| > \omega_p$. These electromagnetic modes
are superluminal and approaches the 
electromagnetic plasma modes described by $\omega^2 = \omega_p^2 + c^2 k^2$, for increasing $k$. However, since these solutions have $\omega /k > c$, 
they undergo no collisionless damping or growing.
Also remark that the dispersion properties of these 
superluminal plasma waves does not change too much in 
the quantum approach, for physically relevant choices of parameters.

\begin{figure}[htb]
\centering
\includegraphics[width=60mm]{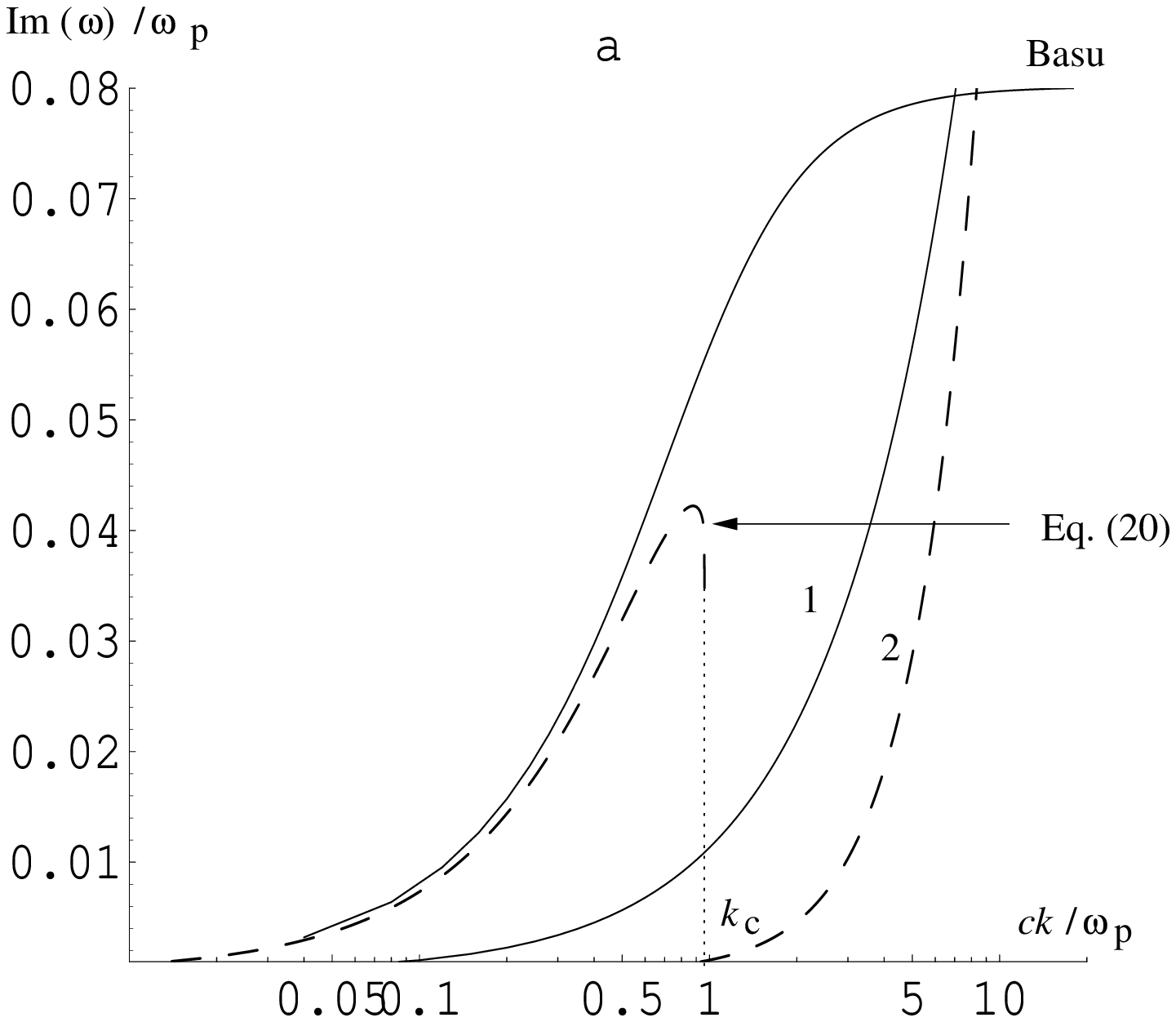}
\includegraphics[width=60mm]{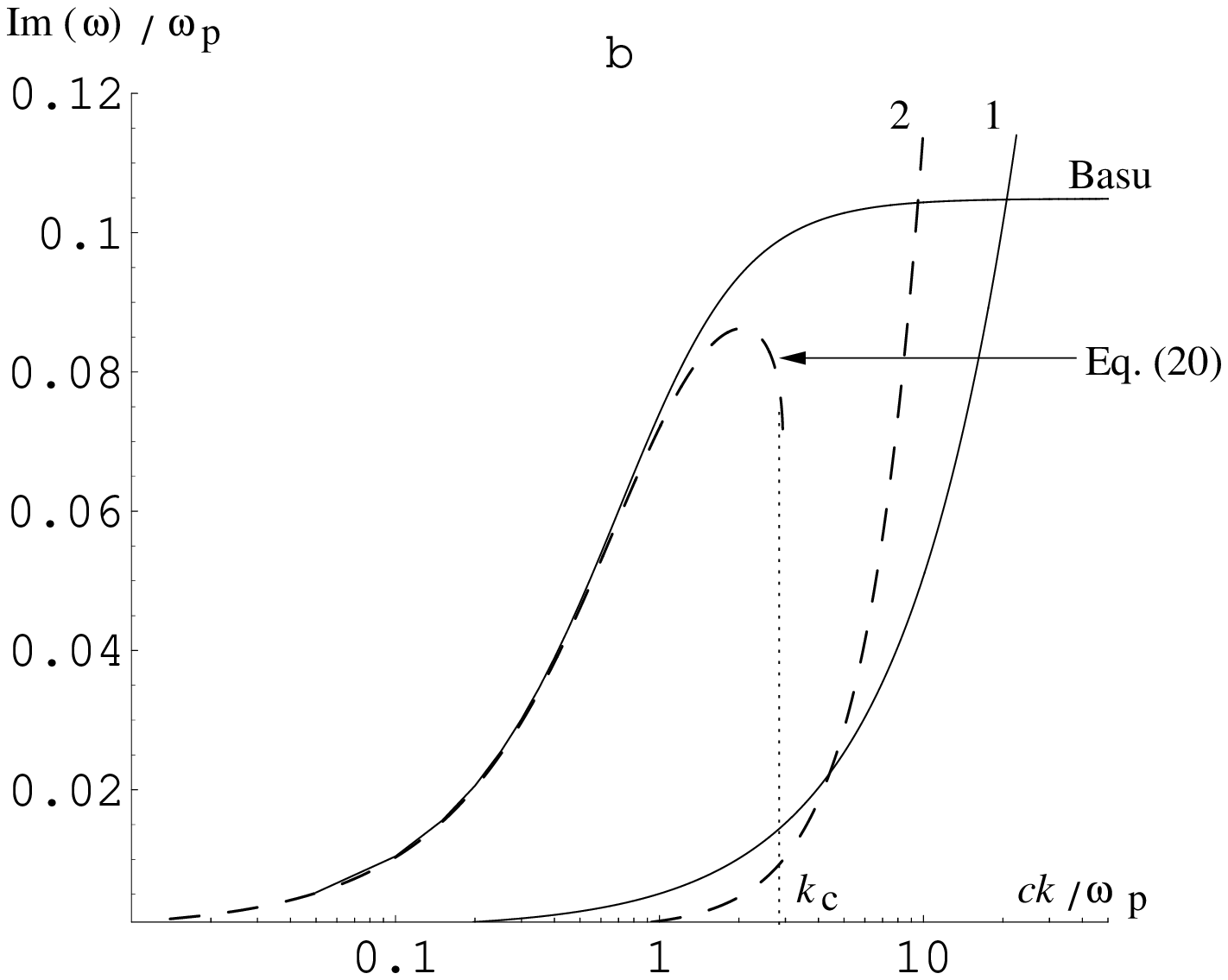}
\caption{With the solid lines 1 and 2, are shown the Weibel growth rates 
obtained from the Eqs. (\ref{e13}) and (\ref{e20}), respectively, 
for two temperature anisotropies (a) $T_{\perp} /T_{\parallel} = 100$ 
and (b) $T_{\perp} /T_{\parallel} = 500$.
The aperiodic solutions of Eq. (\ref{e20}) are even more limited 
to wave numbers $k<k_{c,2} < k_{c,1}$ by the quantum effects 
of the second kind.} \label{fig3} 
\end{figure}

In (\ref{e13}), the quantum corrections of the first kind are proportional to $T_\bot$ and hence to temperature anisotropy. This is because the nature of this modification comes from wave-packet spreading. In an opposite way, in (\ref{e20}) the quantum corrections of the second kind are connected to wave-packet overlap, since, for extreme temperature anisotropy, 
\be
\frac{3k^2 T_{\parallel}(T_\bot - T_{\parallel})}{m\omega^2 T_\bot}\frac{{\rm Li}_{7/2}(- e^{\beta\mu})}{{\rm Li}_{3/2}(- e^{\beta\mu})} \simeq \frac{3k^2 T_{\parallel}}{m\omega^2}\frac{{\rm Li}_{7/2}(- e^{\beta\mu})}{{\rm Li}_{3/2}(- e^{\beta\mu})},\nonumber
\ee
which becomes bigger for larger densities. Therefore, this contribution becomes more evident for increasing degeneracy. 

One may look in the same manner to the unstable solutions of Eq. (\ref{e20}), 
\begin{eqnarray}
\label{e26}
\gamma^2 &=& \frac{k^2 T_\bot}{2 m} \left(\frac{c^2k^2}{\omega_p^2} +1\right)^{-1} \times  \\  &\times&
\left\{1 \pm \left[1- (\frac{12 \, T_{\parallel} (T_\bot - T_{\parallel})}{T_\bot^2}\frac{{\rm Li}_{7/2}(- e^{\beta\mu})}{{\rm Li}_{3/2}(- e^{\beta\mu})}   
+ \frac{\hbar^2 k^2 }{m T_\bot}) (\frac{c^2k^2}{\omega_p^2} +1) \right]^{1/2} \right\} \,, \nonumber
\end{eqnarray} 
which characterize the equilibrium of an anisotropic Fermi-Dirac 
distribution by including the quantum effects of the second kind.
The Weibel growth rates are plotted in Fig. \ref{fig3}
with the solid lines ``2", for two very large temperature anisotropies.
For comparison, the Weibel growth rates provided by  
Eq. (\ref{e13}) are also shown and including only the quantum effects of the first kind (solid lines ``1"), and those provided by the classical approach without 
any quantum effects (dashed lines). In this case, the instability
is limited to smaller wave-numbers $k < k_{c,2} < k_{c,1}$, where $k_{c,2}$ follows from (\ref{e26}), 
\begin{equation}
\label{e27}
k_{c,2}^2 = \frac{\omega_{p}^2}{2c^2} \left\{ \left[ \left(1 -\frac{12 mc^2 T_{\parallel}(T_\bot - T_{\parallel})}{\hbar^2 \omega_p^2 T_\bot} \, 
\frac{{\rm Li}_{7/2}(- e^{\beta\mu})}{{\rm Li}_{3/2}(- e^{\beta\mu})} \right)^2 
+ \frac{4 T_\bot m c^2}{\hbar^2 \omega_{p}^2}\right]^{1/2} -1 \right\} \,.
\end{equation}
and referring to $k_{c,1}$ as those wave-numbers defined in (\ref{e24}), arising from the quantum effects of first kind. 

In the above calculations, only a moderately 
degenerate plasma has been considered, with $\beta \mu \sim 5$. Otherwise, for 
a strongly degenerate one, for instance, with $\beta \mu \sim 200$, 
the aperiodic solutions are completely suppressed, except for unrealistic temperature anisotropies ($T_\bot /T_{\parallel} > 500$).  

\section{Conclusion}

The transition from a kinetic to a fluid-like model in the case of the quantum Weibel instability presents more particularities than one could expect at the first sight. Eq. (\ref{e11}), obtained after retaining terms up to the fourth-order moment of the   equilibrium and perturbed Wigner functions, offers the best way to understand these subtleties. The second term at the right-hand side of (\ref{e11}) is a dispersive term which is always present and is reminiscent of the Bohm potential term at the quantum hydrodynamic model \cite{manfredi}. Since it is universal, here it was identified as a quantum effect of first kind. However, for sufficiently large densities, the term proportional to $I$ in (\ref{e11}) can become the dominant quantum influence, as made clear in Section III. In fact, the cutoff wave-number for instability becomes much smaller for increasing degeneracy reflected in these quantum effects of the second kind. The present work can be relevant not only for applications of Weibel-like instabilities in quantum plasmas as in intense laser-solid interaction experiments but also as a step towards a better conceptual understanding about the origin of the Bohm potential in quantum plasma fluid models, as well as about the transition from kinetic to fluid descriptions for quantum plasmas.  

\vskip .5cm
{\bf Acknowledgments}
\vskip .5cm

The authors acknowledge financial support from the Alexander von Humboldt Foundation. We also thank Prof. Bengt Eliasson for useful discussions.

\end{document}